\documentclass[%
 reprint,
superscriptaddress,
 amsmath,amssymb,
 aps,
prl,
floatfix,
]{revtex4-2}

\usepackage{graphicx}
\usepackage{dcolumn}
\usepackage{bm}
\usepackage{hyperref}

\usepackage{multirow}
\def\degree{\kern-.2em\r{}\kern-.3em}

\begin{document}


\title{\textbf{Collinear, incommensurate antiferromagnetism in  van der Waals magnet $\alpha$-UTe$_3$} 
}%

\author{Hironori Sakai}
\email{sakai.hironori@jaea.go.jp}
\affiliation{Advanced Science Research Center, Japan Atomic Energy Agency, Tokai, Ibaraki 319-1195, Japan}

\author{Chihiro Tabata}%
\affiliation{
 Materials Sciences Research Center, Japan Atomic Energy Agency, Tokai, Ibaraki 319-1195, Japan
}

\author{Koji Kaneko}%
\affiliation{
 Materials Sciences Research Center, Japan Atomic Energy Agency, Tokai, Ibaraki 319-1195, Japan
}

\author{Yoshifumi Tokiwa}
\affiliation{Advanced Science Research Center, Japan Atomic Energy Agency, Tokai, Ibaraki 319-1195, Japan}

\author{Takafumi Kitazawa}
\affiliation{Advanced Science Research Center, Japan Atomic Energy Agency, Tokai, Ibaraki 319-1195, Japan}

\author{Shinsaku Kambe}
\affiliation{Advanced Science Research Center, Japan Atomic Energy Agency, Tokai, Ibaraki 319-1195, Japan}

\author{Yo Tokunaga}
\affiliation{Advanced Science Research Center, Japan Atomic Energy Agency, Tokai, Ibaraki 319-1195, Japan}

\author{Yoshinori Haga}
\affiliation{Advanced Science Research Center, Japan Atomic Energy Agency, Tokai, Ibaraki 319-1195, Japan}

\date{October 13, 2025}

\begin{abstract}
$\alpha$-UTe$_3$, a van der Waals (vdW) actinide compound with a monoclinic ZrSe$_3$-type structure, is a narrow-gap semiconductor with $5f$ moments.
$^{125}$Te NMR reveals strongly anisotropic, layer-confined spin fluctuations below $\sim$20 K, with the $a$-axis component enhanced, and a signal wipeout at the antiferromagnetic (AFM) transition $T_{\rm N}=5$ K.
Single crystal neutron diffraction finds $\bm{q}\approx(0.17,0.5,0)$ and a longitudinal sinusoidal modulation of $a$-axis moments (amplitude $\approx0.8\,\mu_{\rm B}$) with AFM stacking along $b$.
A CEF singlet–singlet induced-moment framework accounts for the easy-axis anisotropy, the small heat-capacity anomaly at $T_{\rm N}$, the reduced ordered moment, and the exchange-driven selection of $\bm{q}$ in this localized $5f$ vdW magnet, establishing a constrained exchange geometry stabilizing this in-plane incommensurate state.
\end{abstract}

\maketitle



Magnetic order in low-dimensional $f$-electron materials is a central theme in quantum matter, where strong spin-orbit coupling and correlations stabilize unconventional states \cite{Tsunetsugu1997The-ground-stat,SiQ:P247:2010,Coleman2010Frustration-and}.
Van der Waals (vdW) crystals offer a clean route to two-dimensional (2D) materials: weak interlayer bonding suppresses interlayer exchange, enhances magnetic anisotropy, and yields cleavable, electronically active layers \cite{Burch2018Magnetism-in-tw, Li2019Intrinsic-Van-D, Yang2021van-der-Waals-M, Park20252D-van-der-Waal}.
However, vdW compounds that truly host strongly correlated $f$ electrons, particularly actinides, remain scarce because of synthesis and stability constraints.

Rare-earth tri-tellurides $R$Te$_3$ ($R$ = Y, La–Nd, Sm, Gd–Tm) \cite{Ru2008Magnetic-proper, Ru2008Effect-of-chemi, Ma2025Thermoelectric-}, crystallizing in the orthorhombic NdTe$_3$-type structure, are metallic prototypes of incommensurate charge density wave (CDW) order with localized $4f$ magnetism.
On the actinide side, orthorhombic $\beta$-UTe$_3$ lies on the localized $5f$ side but in a Kondo-coherent metallic regime that supports 2D ferromagnetism \cite{Thomas2025Enhanced-two-di}.
By contrast, monoclinic $\alpha$-UTe$_3$ is a cleavable ZrSe$_3$-type compound whose basic properties remain largely unexplored.

In this Letter, single-crystal measurements establish that monoclinic $\alpha$-UTe$_3$ is a narrow-gap semiconductor \cite{Blaise1981The-electrical-} and resolve prior inconsistencies regarding its magnetic transition \cite{Janus1982Magnetic-proper,NOEL1986265} by demonstrating antiferromagnetic (AFM) order at $T_{\rm N}=5$~K.
Microscopic probes reveal layer-confined spin anisotropy and a collinear, amplitude-modulated incommensurate AFM state.
Strikingly for a localized $5f$ magnet, the ordered moment modulates in amplitude rather than direction, a hallmark typically associated with itinerant density-wave systems. We attribute this to a crystal electric field (CEF) singlet–singlet induced-moment mechanism, in which exchange admixes two singlets and generates a Van Vleck background, stabilized by exchange pathways constrained by the vdW structure.


\begin{figure}[tbp]
\includegraphics[width=8cm]{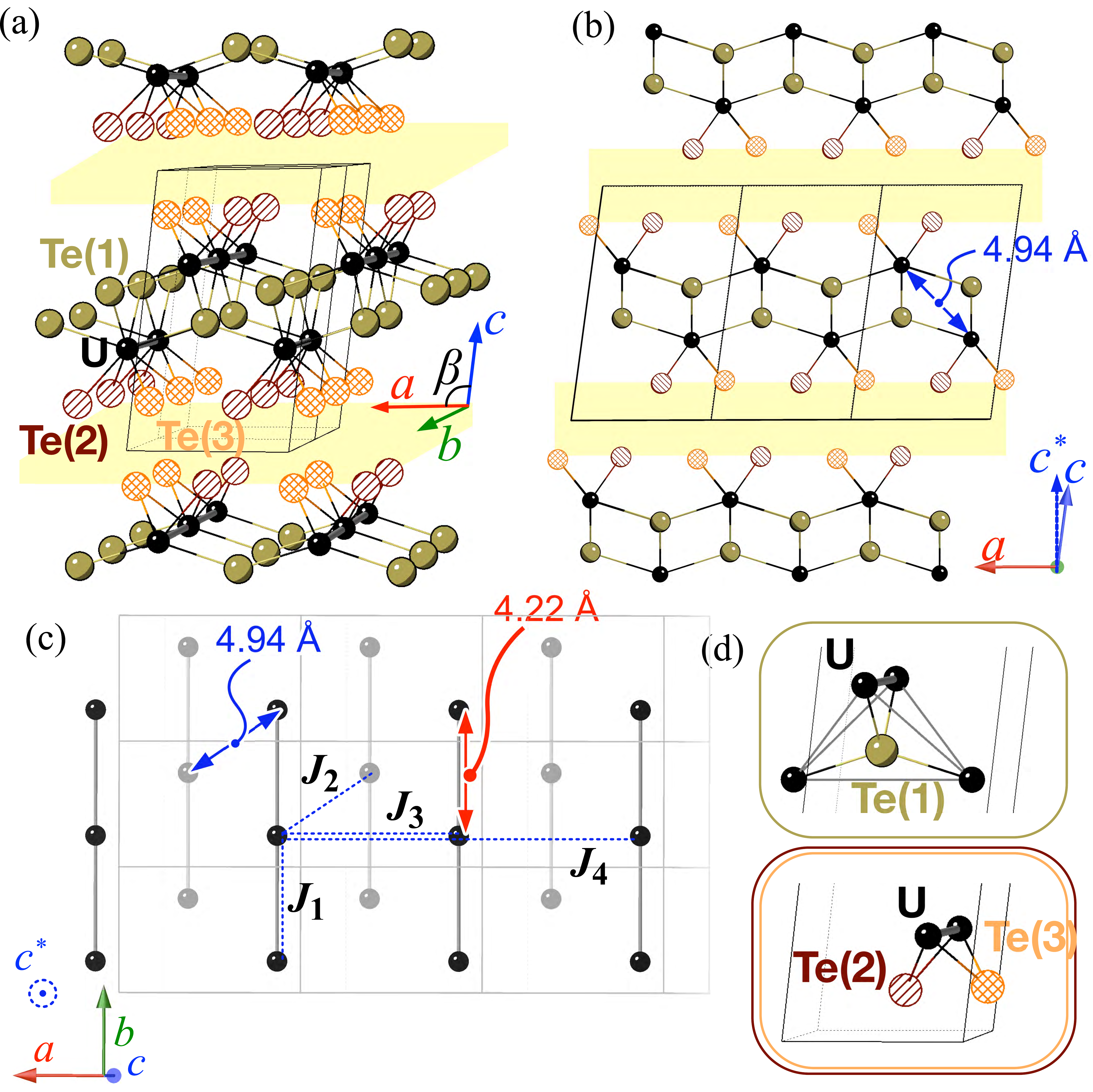}
\caption{\label{fig:crystalstructure} (a) Crystal structure of $\alpha$-UTe$_3$, highlighting van der Waals spacings (yellow bands).
(b) Projection onto the $ac$ ($c^\ast$) plane, viewed along the $b$-axis.
(c) Projection onto the $ab$ plane, showing the one-dimensional U chains along $b$. The $J_i$ denote exchange interactions between U ions.
(d) Local environments of the three crystallographically distinct Te sites: Te(1), Te(2), and Te(3).}
\end{figure}


Single crystals of $\alpha$-UTe$_3$ were grown using the chemical vapor transport method with iodine as the transport agent.
Single-crystal X-ray diffraction confirms the monoclinic ZrSe$_3$-type structure, consistent with prior reports \cite{Boehme1992An-investigatio,Stowe1997Uncommon-valenc,Patschke2001CuxUTe3:--Stabi}.
The refined lattice parameters are $a=6.1047(2)$~\AA, $b=4.2245(15)$~\AA, $c=10.3202(4)$~\AA, and $\beta=97.869(6)^\circ$.
The refined atomic positions identify the crystals as the $\alpha$ phase [Fig.~\ref{fig:crystalstructure}; Table~\ref{tab:alphaUTe3}, End Matter].
In this structure, Te(2) and Te(3) define vdW spacings, while U and Te(1) form the electronically active layers, yielding a soft lattice that cleaves easily along the $ab$ plane.
The shortest U--U distance is $\sim$4.2~\AA\ and the next-nearest is $\sim$4.9~\AA, both exceeding the Hill limit ($\sim$3.5~\AA) \cite{Hill1970Early-actinides} and indicating predominantly localized $5f$ electrons.
The uranium atoms forms a corrugated network of distorted triangles, producing a one-dimensional alignment along the $b$ axis.

\begin{figure}[tbp]
\includegraphics[width=8.5cm]{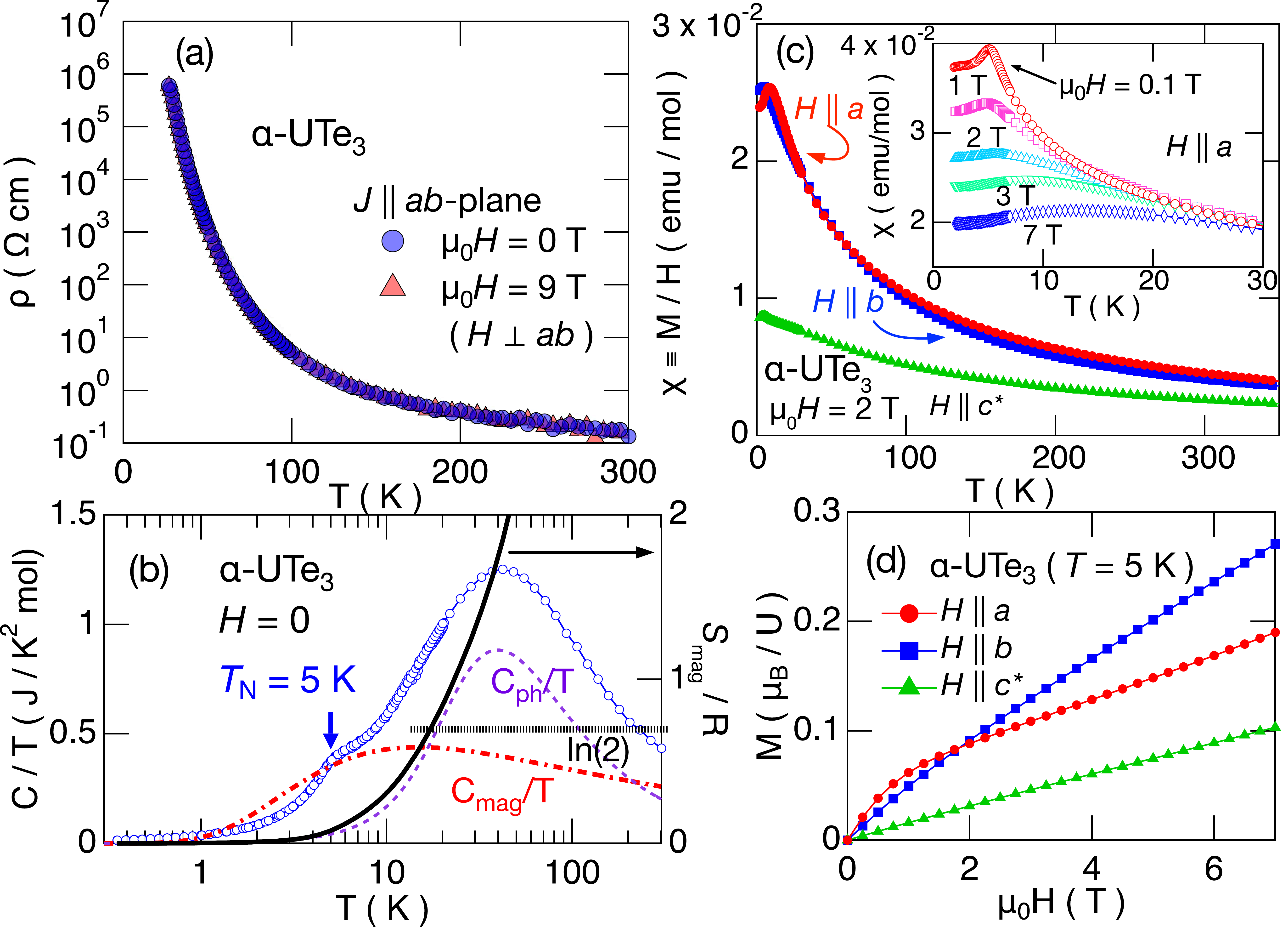}
\caption{\label{fig:rhoT_CT_chiT} (a) Temperature dependence of electrical resistivity $\rho(T)$ for $\alpha$-UTe$_3$ with current in the $ab$-plane under $\mu_0H = 0$ and 9~T ($H \perp ab$).
(b) Specific heat divided by temperature $C/T$ at zero field, with phonon ($C_{\rm ph}/T$), magnetic ($C_{\rm mag}/T$) contributions, and integrated entropy $S_{\rm mag}/R$ (right axis).
(c) Magnetic susceptibility $\chi(T) \equiv M/H$ for $H \parallel a$, $b$, and $c^\ast$; inset: field variation of $\chi(T)$ for $H \parallel a$.
(d) Magnetization $M(H)$ at $T = 5$~K for $H \parallel a$, $b$, and $c^\ast$.}
\end{figure}

On structural-chemistry grounds, the formal uranium valence in $\alpha$-UTe$_3$ is U$^{4+}$ \cite{Stowe1997Uncommon-valenc}, i.e., (U$^{4+}$Te$^{2-}$)(Te$^{-}$)$_2$, by analogy to ZrTe$_3$ \cite{Stowe1998Crystal-Structu}. This stands in clear contrast to rare-earth $R$Te$_3$ compounds \cite{Ru2008Magnetic-proper, Ru2008Effect-of-chemi, Ma2025Thermoelectric-}, which are good metals commonly written as ($R^{3+}$Te$^{2-}$)(Te$^{0.5-}$)$_2$; the Te square net forming vdW spacing host a quasi-2D 5$p$ band \cite{Brouet2008Angle-resolved-}.
This picture is consistent with the temperature ($T$) dependence of the electrical resistivity ($\rho$) for $\alpha$-UTe$_3$, shown in Fig.~\ref{fig:rhoT_CT_chiT}(a).
$\rho(T)$ increases monotonically and approximately exponentially on cooling, consistent with the previous report \cite{Blaise1981The-electrical-}.
The vdW Te layer likely donates charge to the U–Te slabs, accounting for the semiconductivity with a narrow gap.
Below $\sim30$ K, $\rho$ exceeds our measurable range; measurements at lower $T$ require a micro-current setup. From $\rho \propto \exp(\Delta_{\rm g}/k_{\rm B}T)$, we estimate $\Delta_{\rm g}/k_{\rm B}\approx140$ K for $T\lesssim50$ K and $\approx230$ K at higher $T$.
No magnetoresistance is observed up to 9 T, indicating weak coupling between magnetism and charge transport.


Figure~\ref{fig:rhoT_CT_chiT}(b) shows $C/T$ of $\alpha$-UTe$_3$ at zero field.
A kink appears at $T_{\rm N}=5$~K without a sharp $\lambda$ anomaly.
Below $T_{\rm N}$, $C/T$ tends to zero within our resolution, in line with the semiconducting $\rho(T)$.
Subtracting the phonon contribution from a Debye fit with $\Theta_D=142(2)$ K yields $C_{\rm mag}=C-C_{\rm ph}$ and $S_{\rm mag}=\int(C_{\rm mag}/T)\,dT$.
The entropy released at the transition is small, $\Delta S_{\rm mag}(T_{\rm N})\approx0.1\,R\ln2$, while the cumulative $S_{\rm mag}$ approaches $R\ln2$ by $\sim$20~K.
In monoclinic symmetry the $J=4$ multiplet of U$^{4+}$ splits into nine singlets.
Adopting a two-level description for the lowest manifold, the entropy may suggest $\Delta_{\rm CEF}/k_{\rm B}\sim10$~K, but the Schottky peak overlaps with short-range magnetic contributions that smear the maximum and advance $S_{\rm mag}(T)$, biasing a rigid two-level fit toward smaller $\Delta_{\rm CEF}$.
The low-$T$ tail of $C_{\rm mag}(T)$ is described over the fit window by $C_{\rm mag}\propto T^{-\alpha}\exp(-E_{\rm m}/k_{\rm B}T)$ with $\alpha=0.25(3)$ and $E_{\rm m}/k_{\rm B}=3.7(5)$ K, which we regard as an activation scale of low-energy magnetic excitations.
Accounting for these effects, we estimate $\Delta_{\rm CEF}/k_{\rm B}\approx20$ K.
A previous Raman study \cite{Nouvel1987Magnetic-orderi} may indicate a quite rough $\sim80$ K; despite uncertain spectral assignment, the order of magnitude is consistent with our estimate.


Figure~\ref{fig:rhoT_CT_chiT}(c) shows $\chi(T)\equiv M/H$ at $\mu_0H=2$ T for $H\parallel a$, $b$, and $c^\ast$.
Earlier reports gave inconsistent $T_{\rm N}$ values (11~K \cite{Janus1982Magnetic-proper} and 5~K \cite{NOEL1986265}) and anomalously small absolute $\chi$; our data resolve both by establishing $T_{\rm N}=5$ K on a consistent absolute scale.
Above 100~K, $\chi(T)$ follows Curie–Weiss behavior.
Fits for $T\!>\!250$~K give $\mu_{\rm eff}=3.5(2)$, $3.4(2)$, and $2.9(1)$~$\mu_{\rm B}$/U with $\theta_{\rm W}=-45(1)$, $-60(1)$, and $-102(1)$~K for $H\parallel a$, $b$, and $c^\ast$, respectively.
These moments are slightly smaller than the U$^{4+}$ free-ion value (3.58~$\mu_{\rm B}$).
The easy-plane anisotropy reflects the low-symmetry ligand field of Te, and the negative $\theta_{\rm W}$ indicates a dominant antiferromagnetic interaction, likely arising via U–Te–U super-exchange.
The suppression of $T_{\rm N}$ relative to $|\theta_{\rm W}|$ may stem from magnetic frustration and the proximity to a CEF-singlet ground state.

Figure~\ref{fig:rhoT_CT_chiT}(d) displays $M(H)$ at $T=5$~K (just above $T_{\rm N}$).
For $H\!\parallel\!a$, $M(H)$ is concave down, reminiscent of the magnetization process in magnetic semiconductors (e.g., $\beta$-US$_2$ \cite{IkedaS:JPSJ78:2009}), whereas for $H\!\parallel\!b$ and $H\!\parallel\!c$ it is nearly linear; thus $a$ is easier at low fields, but $b$ becomes easier beyond $\sim$2~T.
Consistently, $\chi_a=M_a/H$ shows an apparent low-$T$ enhancement in weak fields [inset of Fig.~\ref{fig:rhoT_CT_chiT}(c)].
The anisotropy between $M_a$ and $M_b$ becomes pronounced below $\sim$20~K, signaling developing spin anisotropy and precursor correlations.
At low fields, $\chi(T)$ exhibits a sharp cusp at $T_{\rm N}$; with increasing field, the cusp shifts to higher $T$ and broadens into a maximum, reminiscent of low-dimensional magnets.


\begin{figure}[!htbp]
\includegraphics[width=8.5cm]{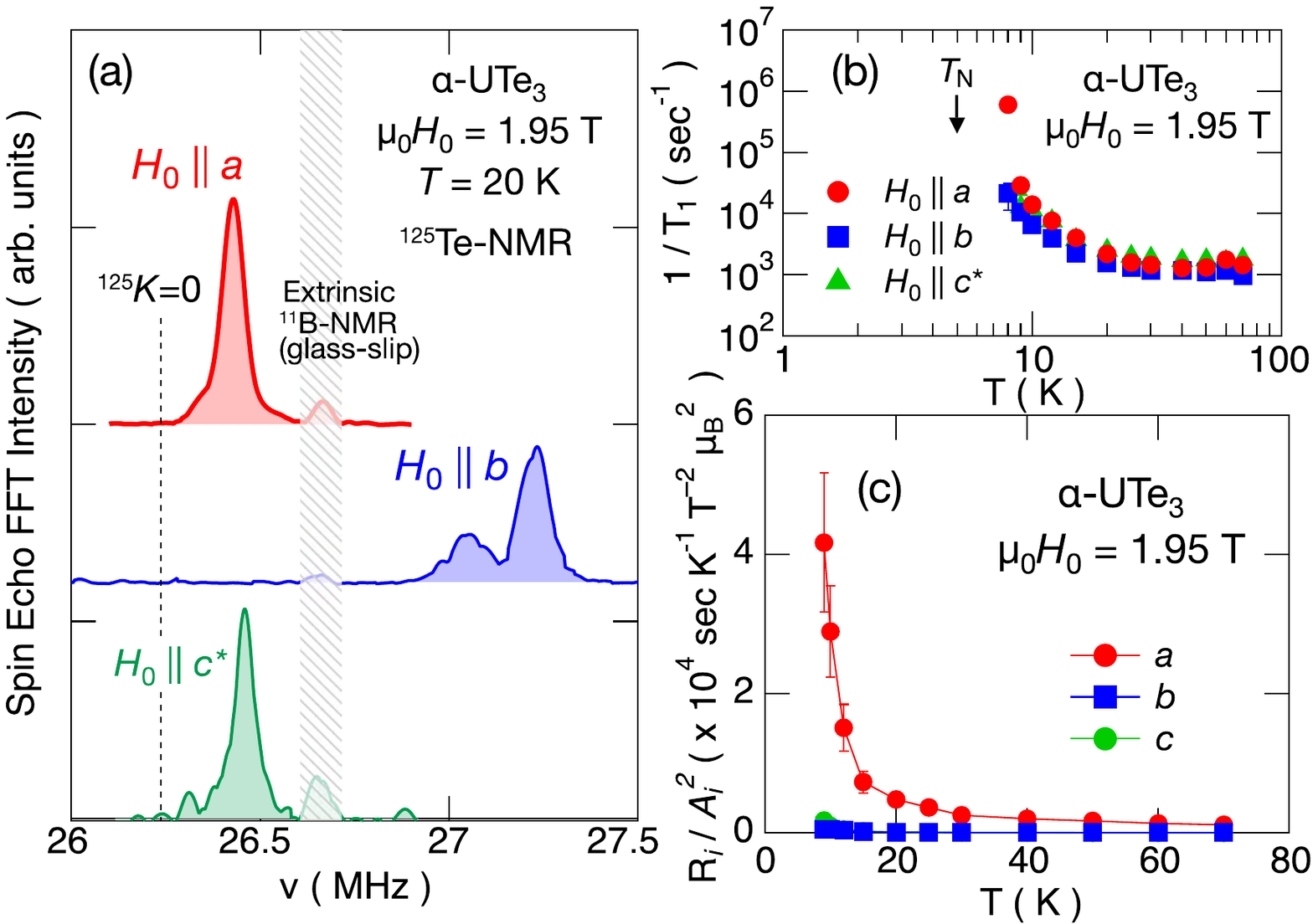}
\caption{\label{fig:NMR} (a) $^{125}$Te-NMR spectra of $\alpha$-UTe$_3$ measured at $T = 20$ K and $\mu_0H_0 = 1.95$ T for magnetic fields applied along the $a$-, $b$-, and $c^\ast$-axes. A small extrinsic $^{11}$B-NMR signal from a glass slip is shaded by a gray-hatched band.
(b) Temperature dependence of the nuclear spin-lattice relaxation rate $1/T_1$ for $H_0 \parallel a$, $b$, and $c^\ast$.
(c) Temperature dependence of the directional spin fluctuation quantity $R_i/|A_i|^2$ for $i = a$, $b$, and $c^\ast$. The definition is given in the main text.}
\end{figure}


Thus, macroscopic measurements establish that $\alpha$-UTe$_3$ is a vdW magnet with localized $5f$ moments, pronounced easy-plane anisotropy.
To gain further microscopic insight into the dynamic magnetic responses, we performed $^{125}$Te nuclear magnetic resonance (NMR) measurements.
Figure~\ref{fig:NMR}(a) shows $^{125}$Te NMR spectra at $T=20$~K and $\mu_0H_0=1.95$~T for $H\!\parallel\!a$, $b$, and $c^\ast$.
Because $^{125}$Te has $I=1/2$, each magnetically inequivalent site gives a single line; with three Te sites [Te(1–3), Fig.~\ref{fig:crystalstructure}] up to three lines are expected.
For $H\!\parallel\!a$ and $H\!\parallel\!c^\ast$ only one peak appears, indicating nearly identical Knight shifts at all Te sites.
For $H\!\parallel\!b$ the spectrum splits into two peaks with an intensity ratio of roughly 1:2.
Based on the local coordination shown in Fig.~\ref{fig:crystalstructure}(c), we assign the weaker peak to Te(1) in the U–Te slabs and the stronger to overlapping Te(2)/Te(3) resonances from the Te–Te bilayers.

Angle-dependent spectra [Fig.~\ref{fig:NMRspectra}, End Matter] did not resolve Te(2) from Te(3), consistent with nearly identical magnetic environments.
We therefore treat the unresolved line as a common Te(2)/Te(3) contribution and determine the Knight shifts $K_i(T)$ for each field orientation (Fig.~\ref{fig:KT_Kchi}(a), End Matter).
From the linear $K_i$–$\chi_i$ relation, $K_i(T)=A_i\chi_i(T)$, we extract hyperfine couplings (Fig.~\ref{fig:KT_Kchi}(b), End Matter): $A_a=0.10(2)$, $A_b=0.91(4)$, and $A_{c^\ast}=0.57(5)$~T/$\mu_{\rm B}$.
The strong anisotropy of $A_i$ exceeds dipolar expectations, indicating anisotropic $5f$–Te hybridization due to direction-dependent U–Te covalency.

We measured the nuclear spin-lattice relaxation rate $1/T_1$ on the Te(2)/Te(3) peak for $H_0\!\parallel\!b$ and on the common Te peak for $H_0\!\parallel\!a$ and $c^\ast$ to probe the anisotropy and $T$ dependence of spin fluctuations [Fig.~\ref{fig:NMR}(b)].
The relaxation rate $1/T_1$ senses spin fluctuations transverse to the field.
In general, for $H_0\!\parallel\!i$ ($i=a,b,c^\ast$),
\begin{equation}
(1/T_1)_{H_0\!\parallel i}=2 (\gamma_{\rm n}A_{\perp}/\gamma_{\rm e}\hbar)^{2}k_{\rm B}T\sum_{\bm q}{\chi''_{\perp\!i}(\bm q,\omega_0)/\omega_0},
\label{eq:invT1}
\end{equation}
where $A_{\perp\!i}$ is the hyperfine coupling constant, $\chi''_{\perp \!i}$ is the imaginary part of the dynamical susceptibility, the subscript $\perp\!i$ means component perpendicular to $H_0$), and $\omega_0$ is the Larmor angular frequency \cite{Moriya_LocalMomentLimit}.
Above $\sim20$~K, $1/T_1$ is nearly $T$-independent, consistent with paramagnetic fluctuations of localized U moments.
Below $\sim20$ K, $1/T_1$ undergoes a qualitative change: it surges by more than three decades as $T$ approaches $T_{\rm N}$, while the NMR echo is progressively wiped out. Very close to $T_{\rm N}$, $1/T_1$ exceeds the experimental time window, directly evidencing critical slowing down of low-frequency AFM fluctuations.

To identify which spin components drive this growth, we use the transverse projections based on Eq.~(\ref{eq:invT1}).
For $H_0\!\parallel\!a$, $(T_1T)^{-1}=R_b+R_{c^\ast}$; for $H_0\!\parallel\!b$, $(T_1T)^{-1}=R_a+R_{c^\ast}$; and for $H_0\!\parallel\!c^\ast$, $(T_1T)^{-1}=R_a+R_b$, where we define the direction-resolved fluctuation rate
$R_i\equiv (\gamma_{\rm n}/\gamma_{\rm e}\hbar)^{2}k_{\rm B}\cdot A_i^2\sum_{\bm{q}}{\chi_i''(\bm{q},\omega_0)/\omega_0}$.
Inverting the three projection relations yields the hyperfine-normalized amplitudes $R_i/A_i^2$.
Applying this decomposition [Fig.~\ref{fig:NMR}(c)] reveals a clear and striking trend: below $\sim20$~K the AFM fluctuations become strongly anisotropic, with the $a$-polarized component $R_a/A_a^2$ rising far more rapidly than those along the $b$ and $c^\ast$ axes.
This anisotropic critical growth provides microscopic evidence that the AFM state has a pronounced easy-axis character along $a$.


\begin{figure*}[!htbp]
\includegraphics[width=18cm]{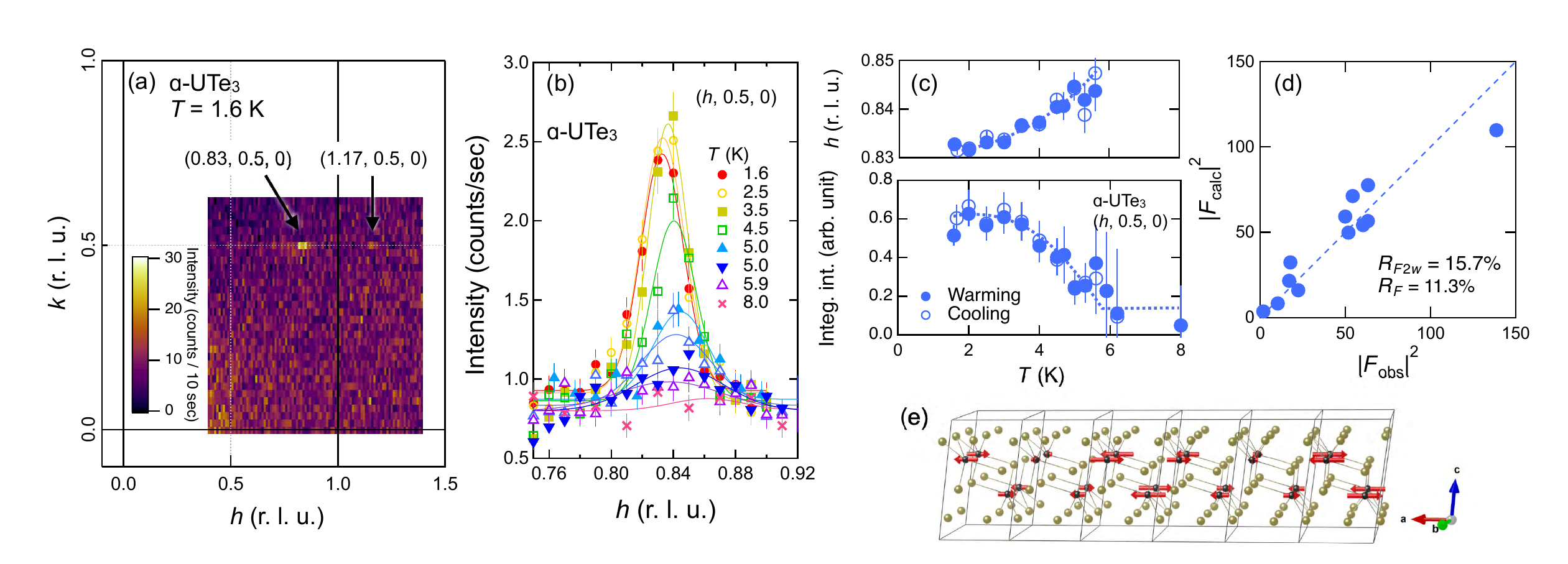}
\caption{\label{fig:ND}
(a) Neutron diffraction intensity map in the $(h, k, 0)$ plane for a single crystal of $\alpha$-UTe$_3$ at $T = 1.6$~K. 
(b) Temperature dependence of the peak profile of the magnetic reflection at $\bm{Q}=(\sim 0.83, 0.5, 0)$. Solid lines represent Gaussian fits.
(c) Peak positions and integrated intensity of the magnetic reflection at $\bm{Q}=(\sim 0.83, 0.5, 0)$. The dotted lines are guides to the eye.
(d) $|F_{\rm calc}|^2$ vs $|F_{\rm obs}|^2$ plot of the magnetic reflections observed at 1.6~K.
The dashed line is a guide to the eye.
(e) The magnetic structure model employed in the fitting analysis of the neutron data. The illustration was generated using VESTA\cite{vesta}.
}
\end{figure*}

Since the rapid increase of the relaxation rate rendered NMR signals unobservable in the AFM state, we performed single-crystal neutron diffraction experiments to investigate the magnetic order in $\alpha$-UTe$_3$.
A thin, plate-like crystal with its largest face in the $ab$ plane was mounted with the ($h, k, 0$) plane horizontal.
Our initial search for magnetic reflections along high-symmetry axes [100] and [010] was failed to detect any magnetic peaks within the experimental sensitivity.
We then mapped the $(h,k,0)$ plane, as shown in Fig.~\ref{fig:ND}(a) and revealed magnetic reflections at (0.83, 0.5, 0) and (1.17, 0.5, 0).
Figure~\ref{fig:ND}(b) shows the $T$ dependence of the $h$-scan across the magnetic Bragg peak at $(0.83, 0.5, 0)$.
The intensity gradually decreases with increasing $T$, remaining finite up to $\sim$5.3~K and vanishing by 5.9~K, indicating $T_{\rm N}$ lies between these temperatures.
Notably, the anomaly at $T_{\rm N}$ in the specific heat is rather subtle, and the associated entropy change is small, suggesting that short-range magnetic correlations may already develop slightly above $T_{\rm N}$.

Figure~\ref{fig:ND}(c) displays the peak position and integrated intensity of this magnetic peak obtained by Gaussian fitting and plotted as a function of $T$ for both warming and cooling.
The peak position continuously varies with temperature, indicating the incommensurate nature of the magnetic order.
No clear thermal hysteresis is observed across $T_{\rm N}$ within the statistical error, consistent with the magnetization data.

Neutron diffraction unambiguously establishes an incommensurate AFM propagation vector $\bm{q}=(\zeta,0.5,0)$ with $\zeta=0.170(1)$ at 1.6 K. With NMR constraints, least-squares refinement yields U$^{4+}$ moments aligned along $a$ (easy axis) that form a longitudinal sinusoidal order with $\bm{q}\!\parallel\!a$, while alternating stacking along $b$ forms AFM chains.
The model reproduces the measured intensities: in Fig.~\ref{fig:ND}(d) $|F_{\rm calc}|^{2}$ scales approximately linearly with $|F_{\rm o}|^{2}$ despite the limited number of reflections and modest statistics, indicating that the overall intensity distribution is captured.
The real-space arrangement is shown in Fig.~\ref{fig:ND}(e).
The refined modulation amplitude is $0.79(4),\mu_{\rm B}$, well below the U$^{4+}$ free-ion value $g_JJ=3.2\,\mu_{\rm B}$.

What mechanism stabilizes a collinear, longitudinally sinusoidal incommensurate AFM order in semiconductive vdW magnet?
A CDW route as in metallic $R$Te$_3$ \cite{Ru2008Magnetic-proper,Ru2008Effect-of-chemi,Ma2025Thermoelectric-} would require a CDW at or above $T_{\rm N}$ that differentiates U sites and modulates the ordered moment, implying large CEF variations and a lattice distortion.
In magnetic $R$ members the low-$T$ order typically locks to the preexisting CDW; for example, the DyTe$_3$ cycloid encodes incommensurability in helicity rather than amplitude and exhibits a sharp $\lambda$-type specific-heat peak \cite{Akatsuka2024Non-coplanar-he}, none of which we observe.

An alternative consistent with all observations is a CEF singlet–induced-moment state, realized in Pr$_3$Tl \cite{Birgeneau1971Magnetic-Excito,Buyers1975Temperature-dep}, Pr$_3$In \cite{Fanelli2008Magnetic-order-}, PrAu$_2$Si$_2$ \cite{Goremychkin2008Spin-glass-orde}, and in TmSe and TmS \cite{Holtzberg1977Hyperfine-Speci}.
In this scenario a nonmagnetic singlet orders via exchange-induced admixing with a nearby CEF level; a recent unified treatment \cite{Thalmeier2024Induced-quantum} for non-Kramers $f$ ions, applied to $\alpha$-UTe$_3$, reproduces the main thermodynamic and magnetic signatures.

Within this framework \cite{Thalmeier2024Induced-quantum}, a single control parameter $\xi$ governs the instability from a Van Vleck paramagnet to induced-moment order, with the quantum critical point (QCP) at $\xi=1$.
In the singlet–singlet CEF model $\xi=1/\tanh(\Delta_{\rm CEF}/2T_{\rm N})$; using $\Delta_{\rm CEF}/k_{\rm B}\approx20$ K and $T_{\rm N}=5$ K gives $\xi=1.03$, placing the system near the QCP.
This model explains the uniaxial spin orientation through single-ion anisotropy and yields both a tiny specific-heat jump at $T_{\rm N}$ and a reduced ordered moment.

Quantitatively, for $\xi=1.03$ we obtain $\delta C/C^{+}(T_{\rm N})\approx0.14$ (with $C^{+}(T_{\rm N})$ just above $T_{\rm N}$) where $\delta C/C^{+}(T_{\rm N})\to0$ as $\xi\!\to1^{+}$.
The ordered moment at $T\to0$ is $\mu_{\rm ord}=\mu_{01}\sqrt{1-\xi^{-2}}=0.24\,\mu_{01}$, where $\mu_{01}=g_J\mu_{\rm B}|\langle0|J_{\parallel}|1\rangle|$.
The uniform susceptibility in mean field is $\chi(T)=\chi_0(T)/[1-I_e\chi_0(T)]$, with $I_e$ the molecular-field exchange constant, consistent with $E_m=3.7$~K estimated from $C_{\rm mag}$;
this forms Curie-Weiss behavior at high $T$ and fixes $T_{\rm N}$ via $1=I_e\chi_0(T_{\rm N})$.
The dynamical susceptibility within this model also reproduces the sharp increase of $1/T_1$ just above $T_{\rm N}$.

Because $\chi_0(T)$ is effectively $\bm{q}$ independent, the instability generalizes to $1-\chi_0(T_{\rm N})J(\bm{q})=0$ by replacing $I_e\to J(\bm{q})$, so $\bm{Q}=\arg\max_{\bm{q}}J(\bm{q})$.
Along $a$, a minimal kernel $J(h)=-2J_1+A_1\cos(2\pi h)+A_2\cos(4\pi h)$ gives $\cos(2\pi h^{\ast})=-A_1/(4A_2)$, hence $h^{\ast}$ is set by $J_4/J_3$ ($A_1\!\sim\!J_3$, $A_2\!\sim\!J_4$).
Here $J_i$ are the exchanges in Fig.~\ref{fig:crystalstructure}(c).
On the $k$=1/2 plane, $J_2$ cancels, leaving only the ratio $J_4/J_3$ relevant; even if $|J_3|$ and $|J_4|$ are small in absolute value, any finite values suffice to set $h^{\ast}$.
That only $J_4/J_3$ matters, both longer-range interchain couplings in one $c$ plane, underscores the vdW-constrained exchange geometry of $\alpha$-UTe$_3$.

In summary, semiconducting $\alpha$-UTe$_3$ hosts a collinear, $a$-axis longitudinally sinusoidal incommensurate AFM state, explained by a CEF singlet–induced-moment framework.
Within this picture $\alpha$-UTe$_3$ lies near the Van Vleck paramagnetic QCP.
This work further motivates systematic exploration of tuning $\bm{q}$ and $T_{\rm N}$ by pressure or substitution via exchange anisotropy.
More broadly, $\alpha$-UTe$_3$ exemplifies a vdW-constrained pathway where localized $5f$ physics sets $\bm{q}$ and moment polarization, relevant to 2D ferromagnetism in $\beta$-UTe$_3$ \cite{Thomas2025Enhanced-two-di} and to pressure-induced incommensurate AFM order in the spin-triplet superconductor UTe$_2$.

\vspace{1em}
We thank Y. Hirose, P. F. S. Rosa, E. D. Bauer, J. D. Thompson, T. Park, L. Havela, K. Gofryk, D. Kaczorowski, and J-G. Park for fruitful discussions.
This work was also supported by JSPS KAKENHI Grant Nos. 21H04987, 23H04867, 23H04871, 23K03332, 23K25829, 24KK0062, and 24K00590 by the
JAEA REIMEI Research Program.
The experiment on TAS-1 was performed under Proposal Nos. D836 and D1053.

%


\subsection*{Appendix A: Crystal Growth and Characterization Details}
Single crystals of $\alpha$-UTe$_3$ were grown by iodine-assisted chemical vapor transport.
Etched U (brief dilute HNO$_3$), 6N Te, and 5N I$_2$ (loading $\sim0.25$ mg cm$^{-3}$) were loaded stoichiometrically into a quartz ampoule evacuated to $\sim10^{-6}$ mbar.
To prevent I$_2$ loss during pumping, iodine was pre-sealed in a quartz capillary and released in situ by breaking it inside the evacuated ampoule.
Growth used a two-zone gradient of 550–650$^\circ$C for several weeks.
Plate-like crystals a few millimeters across and 0.1–0.3 mm thick were obtained, rinsed with ethanol, vacuum dried, and showed broad $ab$-plane facets consistent with the layered structure.

\begin{table}[!htbp]
\caption{\label{tab:alphaUTe3} Fractional atomic coordinates and equivalent isotropic displacement parameters $B_{\rm eq}$ for $\alpha$-UTe$_3$ refined from single-crystal X-ray diffraction. The final refinement converged with $R_1 = 0.0390$ ($I > 2\sigma$), $wR_2 = 0.1247$.}
\begin{ruledtabular}
\begin{tabular}{lcccc}
Atom  & $x$ & $y$ & $z$ & $B_{\rm eq}$ (\AA$^2$) $^{a}$ \\
\hline
U1    & 0.21482(7) & 0.250000 & 0.34127(4) & 0.0046(3)  \\
Te1   & 0.26509(13) & 0.750000 & 0.56456(8) & 0.0049(3)  \\
Te2   & 0.39143(15) & 0.750000 & 0.15690(9) & 0.0111(3)  \\
Te3   & -0.05887(15) & 0.750000 & 0.16728(9) & 0.0102(3)  \\
\end{tabular}
\end{ruledtabular}
\begin{flushleft}
$^a$ Equivalent isotropic atomic displacement parameter is defined as 
$B_{\rm eq} = (8\pi^2/3) [ U_{11}(aa^*)^2 + U_{22}(bb^*)^2 + U_{33}(cc^*)^2 ]$ where $U_{ij}$ is experimentally obtained anisotropic atomic displacement tensor.
\end{flushleft}
\end{table}

The crystallographic structure of the samples was determined at room temperature using single-crystal X-ray diffraction (XRD) with graphite-monochromated Mo $K_{\alpha}$ radiation.
The scattered X-ray beam was recorded on an imaging plate detector (R-AXIS RAPID, Rigaku Corp., Tokyo, Japan).
An empirical absorption correction was applied before structural analysis. Structural solutions and refinement of crystallographic parameters were performed using the SHELX software suite \cite{Sheldrick2015Crystal-structu}.
Electron probe microanalysis (EPMA) was conducted with wavelength-dispersive spectrometers installed in a scanning electron microscope (JXA-iHP200F, JEOL Ltd., Tokyo, Japan).

$\alpha$-UTe$_3$ crystallizes in a monoclinic ZrSe$_3$-type structure with space group $P2_1/m$.
The refined lattice parameters are $a=6.1047(2)$~\AA, $b=4.2245(15)$~\AA, $c=10.3202(4)$~\AA, and $\beta=97.869(6)^\circ$.
The atomic coordinates and equivalent isotropic displacement parameters obtained from the refinement are listed in Table~\ref{tab:alphaUTe3}.
Electron probe microanalysis confirmed the stoichiometric composition of UTe$_3$.

Electrical resistivity and heat capacity were measured in a PPMS (DynaCool-9, Quantum Design).
Resistivity used an ac four-probe at 190 Hz with currents 10 nA–1 mA for $\mu_0H=0$–9 T down to 0.35 K; contacts were 25-$\mu$m Au wires spot-welded to a flat facet.
Heat capacity employed the relaxation two-$\tau$ method over 0.35–300 K.
Magnetization was measured in a SQUID magnetometer (MPMS XL7, Quantum Design) over 2–300 K for $\mu_0H=0.01$–7 T.

\subsection*{Appendix B: Details of NMR experiment}
\begin{figure}[!tb]
\includegraphics[width=8.5cm]{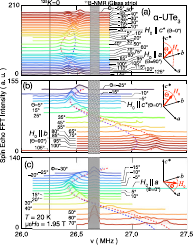}
\caption{\label{fig:NMRspectra}
Field-angle dependence of $^{125}$Te NMR spectra in $\alpha$-UTe$_3$ measured at $T = 20$~K and $\mu_0H = 1.95$~T.
(a) Field rotated in the plane perpendicular to the $b$ axis; angle $\Theta$ is measured from the $c^\ast$ axis.
(b) Field rotated in the plane perpendicular to the $a$ axis; $\Theta$ is defined similarly.
(c) Field rotated in the $ab$ plane (perpendicular to the $c^\ast$ axis); angle $\Phi$ is measured from the $a$ axis.
The vertical line indicates the reference frequency corresponding to $^{125}K = 0$. The hatched area represents an insensitive region due to $^{11}$B NMR signals from the glass strip used to fix the sample.}
\end{figure}

\begin{figure}[!tb]
\includegraphics[width=8.5cm]{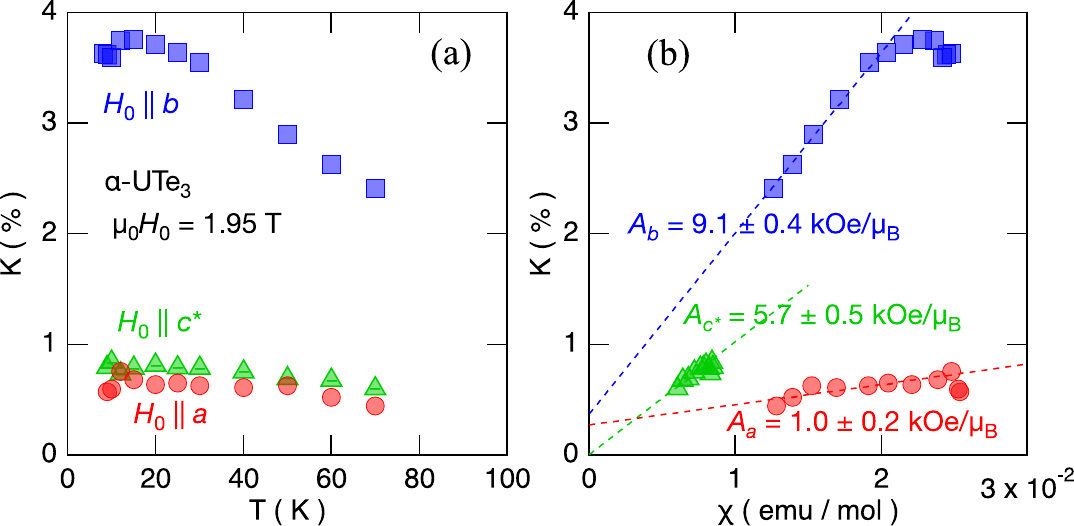}
\caption{\label{fig:KT_Kchi}
(a) Temperature dependence of the Knight shift for the Te(2) and Te(3) sites with the magnetic field applied along the $a$-, $b$-, and $c^\ast$-axes at $\mu_0 H_0 = 1.95$T.
(b) Knight shift $K$ versus magnetic susceptibility $\chi$ ($\chi$ measured at $\mu_0 H_0 = 2$~T).}
\end{figure}

NMR was performed in a cryogen-free 14 T magnet (Cryogenic) over 1.7–100 K using a two-axis goniometer probe and a custom coherent-pulse spectrometer (Thamway).
Frequency-swept spectra were collected by stepwise auto-tuning at fixed field.

Because $^{125}$Te has $I=1/2$, each Te site yields one line; at most three lines are expected from Te(1)–Te(3).
Angular-dependent spectra [Fig.~\ref{fig:NMRspectra}] show accidental coalescence into a single line for $H\!\parallel\!a$ and $H\!\parallel\!c^\ast$, where the Knight shifts of Te(1)–Te(3) coincide.
Tilting the field away resolves two lines: one from Te(1) and a composite Te(2)/Te(3) signal.
For $H\!\parallel\!b$, the spectrum displays a clear 1:2 intensity ratio, consistent with site multiplicity, and all sites exhibit large, anisotropic Knight shifts.

The temperature dependence of the Knight shift for $H\!\parallel\!a$, $b$, and $c^\ast$, determined from the Te(2)/Te(3) resonance positions, is shown in Fig.~\ref{fig:KT_Kchi}(a).
Hyperfine couplings are obtained from the $K$–$\chi$ analysis in Fig.~\ref{fig:KT_Kchi}(b).
Notably, for $H\!\parallel\!a$ the $K$–$\chi$ plot bends downward at low $T$, reducing the apparent slope (effective $A_a$), which may be expected for a CEF–singlet ground state.

Spin-lattice relaxation $1/T_1$ was measured on the Te(2)/Te(3) NMR line by inversion recovery.
Recoveries were fitted to a stretched exponential, $\{M_{\rm n}(\infty)-M_{\rm n}(t)\}/M_{\rm n}(\infty)\propto\exp[-(t/T_1)^\beta]$ with $0<\beta\le1$.
Here $\beta<1$ reflects a distribution of $1/T_1$ arising from overlap of multiple Te sites.
For $H\!\parallel\!b$ and $H\!\parallel\!c^\ast$, $\beta\approx1$, indicating nearly homogeneous relaxation across Te(1)–Te(3).
In contrast, for $H\!\parallel\!a$, $\beta\approx0.5$, implying a distinct $1/T_1$ component from Te(1).

\subsection*{Appendix C: Details of Neutron Diffraction Experiment}
\begin{figure}[!tb]
\includegraphics[width=8.5cm]{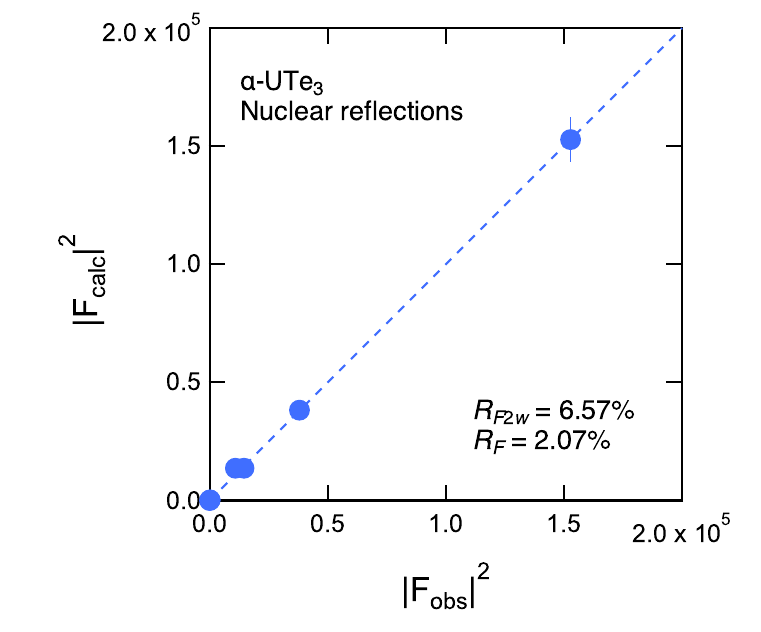}
\caption{\label{fig:Fo-Fc_nuc}
$|F_{\rm calc}|^2$ vs $|F_{\rm obs}|^2$ plot of the nuclear reflections observed at 1.6~K.
The dashed line is a guide to the eye.}
\end{figure}
The neutron diffraction experiments were conducted on the thermal triple-axis spectrometer TAS-1, located at the research reactor JRR-3 of the Japan Atomic Energy Agency in Tokai, Japan.
The spectrometer was configured for elastic scatterings in triple-axis mode with a fixed neutron wavelength of 2.359 \AA, using a vertical-focusing pyrolytic graphite (PG) monochromator and analyzer.
The collimation sequence used was open–80’–40’–80’, and a PG filter was placed before the sample to suppress higher-order contamination.
The single-crystal sample was mounted on an aluminum plate to align the $(hk0)$ horizontal scattering plane. The sample was sealed in an aluminum can filled with helium exchange gas and attached to a cold finger of a dry-type $^4$He top-load cryostat \cite{Kaneko2024CLVTC}, cooling the sample to 1.6 K.

The least-square fitting of data to model crystal and magnetic structures were carried out using the FULLPROF package \cite{FullProf}. 
Before fitting the magnetic reflection data, nuclear reflections were fitted to determine the scale factor.
During the refinement, the atomic displacement parameters were fixed to zero as the measurement temperature was 1.6~K.
The $z$-parameters of atomic coordinates were fixed to values determined in the XRD analysis because of the lack of data of $hkl$ reflections with non-zero $l$, as the scattering plane was $(h, k, 0)$.
The fitting yielded a good agreement with the data and the structure model, as shown in the  $|F_{\rm calc}|^2$ vs $|F_{\rm obs}|^2$ plot in Fig.~\ref{fig:Fo-Fc_nuc}.
\end{document}